\documentclass{amsart}

\usepackage[centertags]{amsmath}
\usepackage{amsfonts}
\usepackage{amssymb}
\usepackage{euscript}
\usepackage{newlfont}
\usepackage{euscript}
\usepackage[colorlinks=true]{hyperref}
\usepackage{amsrefs}
\usepackage{slashed}

\newtheorem{lemma}{Lemma}
\newtheorem{corollary}{Corollary}
\newtheorem{theorem}{Theorem}

\newcommand{\R}{\mathbb R}

\newcommand{\D}{\partial}

\renewcommand{\div}{\operatorname{div}}

\newcommand{\V}{\widetilde{V}}
\newcommand{\U}{\widetilde{U}}
\newcommand{\hatU}{\widehat{U}}
\newcommand{\hatV}{\widehat{V}}

\newcommand{\RN}{Reissner-Nordstr\"om}
\newcommand{\MP}{Majumdar-Papapetrou}

\begin{document}
\title[The Riemannian Penrose Inequality with Charge]{The Riemannian Penrose Inequality with Charge\\ for Multiple
Black Holes}

\author{Marcus Khuri, Gilbert Weinstein \and Sumio Yamada}
\thanks{M.~Khuri was supported by NSF grants DMS-1007156 \& DMS-1308753.}
\address{Department of Mathematics\\ Stony Brook University\\ Stony Brook, NY 11794}
\email{khuri@math.sunysb.edu}
\address{Physics Dept. and Dept.\ of  Computer Science and Mathematics\\ Ariel University of Samaria\\ Ariel 40700,
Israel}
\email{gilbert.weinstein@gmail.com}
\thanks{S.~Yamada by JSPS Grants 23654061 \& 24340009.}
\address{Department of Mathematics\\Gakushuin University\\ Tokyo 171-8588, Japan}
\email{yamada@math.gakushuin.ac.jp}

\begin{abstract}
We present the outline of a proof of the Riemannian Penrose inequality with charge $r\leq m + \sqrt{m^2-q^2}$,
where $A=4\pi r^2$ is the area of the outermost apparent horizon with possibly multiple connected components, $m$ is the
total ADM mass, and $q$ the total
charge of a strongly asymptotically flat initial data set for the Einstein-Maxwell equations, satisfying the
charged dominant energy condition, with no charged matter outside the horizon.
\end{abstract}


\maketitle

\section{Introduction}

In a seminal paper~\cite{penrose1973}, R.~Penrose examined the validity of the cosmic censorhip
conjecture, and outlined a heuristic argument which shows how using also Hawking's area theorem~\cite{hawkingellis}, it
implies a related inequality. In~\cite{penrose1982}, he generalized this heuristic
argument leading to an inequality now referred to as the Penrose inequality. Consider a strongly asymptotically flat
(SAF) Cauchy surface in a spacetime satisfying
the dominant energy condition (DEC), with ADM mass $m$ containing an event horizon of area $A=4\pi r^2$, which undergoes
gravitational collapse and settles to a Kerr solution. Since the ADM mass $m_\infty$ of the final state is no greater
than $m$, and since the area radius $r_\infty$ is no less than $r$, and since for the final state we must have
$m_\infty\geq \frac12 r_\infty$ in order to avoid naked singularities, we must have had $m\geq \frac12 r$ also at the
beginning of the evolution. The event horizon is indiscernible in the original slice without knowing the full evolution.
However, one may replace the event horizon by the outermost minimal area enclosure of the 
apparent horizon, the boundary
of the region admitting trapped surfaces, and obtain the same inequality. A counterexample 
to the Penrose inequality
would therefore have suggested data which leads under the Einstein evolution to naked singularities, while a proof of
the inequality could be viewed as evidence in support of cosmic censorship.

The inequality further simplifies in the time-symmetric case, where the apparent horizon coincides with the outermost
minimal area enclosure. The dominant energy condition reduces now to non-negative scalar curvature of
the Cauchy hypersurface, leading to the Riemannian version of the inequality: the ADM mass and the area radius of the
outermost compact minimal surface in a SAF 3-manifold of non-negative scalar curvature satisfy
$m\geq \frac r2$
with equality if and only if the
manifold is a Schwarzschild slice. Note that this characterizes the Schwarzschild slice as the unique minimizer of
$m$ among all such 3-manifolds admitting an outermost horizon of area $A=4\pi r^2$.

This inequality was first proved in the special case where the horizon is connected by
Huisken--Ilmanen~\cite{huiskenilmanen2001} using the inverse mean curvature flow, an approach proposed by
Jang--Wald~\cite{waldjang}, following Geroch~\cite{geroch1973}
who had shown that the Hawking mass is
non-decreasing under this flow. The inequality was proven in full generality by Bray~\cite{bray2001} using a conformal
flow of the initial Riemannian metric, and the positive mass theorem~\cites{schoenyau1979,witten1981}.

We now turn to the charged case which is slightly more subtle. It is natural to conjecture as above that
the \RN\ spacetime (RN), the charged analog of Schwarzschild spacetime gives rise to the unique minimizer of $m$, given
$r$ and $q$. Since RN satisfies $m=\frac12(r+q^2/r)$ where $q$ is the total charge, one is thus lead to conjecture that
in any SAF data satisfying $R_g\geq 2(|E|^2 + |B|^2)$, where $E$ and $B$ are respectively the electric and magnetic
field, and $R_g$ is the scalar curvature of $g$, we have

\begin{equation} \label{charged-penrose-inequality}
  m \geq \frac12 \left( r + \frac{q^2}{r} \right)
\end{equation}
with equality if and only if the initial data is RN. This is shown
in~\cite{huiskenilmanen2001}, based on Jang~\cite{jang1979}, but only for a connected horizon, since the proof is
based on inverse mean curvature flow. In
fact~\eqref{charged-penrose-inequality} can fail if the horizon is not connected, and a 
counterexample based on \MP\
(MP) initial data with two black holes was constructed in~\cite{weinsteinyamada}.
This counterexample, however, does not suggest a counterexample to cosmic censorhip. This 
is because the
right-hand side of~\eqref{charged-penrose-inequality} is not monotonically increasing in $r$. Indeed, already Jang
observed
that~\eqref{charged-penrose-inequality} is equivalent to two inequalities:
\begin{equation} \label{upper-lower-bound}
  m - \sqrt{m^2 - q^2} \leq r \leq m + \sqrt{m^2-q^2}.
\end{equation}
Cosmic censorship suggests the upper bound always holds, while the counterexample 
in~\cite{weinsteinyamada} violates
the
lower bound.

In this paper, we prove the upper bound in~\eqref{upper-lower-bound} for multiple black holes. By the positive mass
theorem with charge we have $m\geq
|q|$ with equality if and only if the data is MP~\cite{gibbonshawkinghorowitzperry}. Hence if $r \leq |q|$, the
upper
bound in~\eqref{upper-lower-bound} follows immediately
\[
  r\leq |q| \leq m \leq m + \sqrt{m^2-q^2}.
\]
It thus only remains to prove the upper bound under the additional hypothesis $|q| \leq r$. Under this hypothesis, it
is the lower bound that follows immediately
\[
  m \leq |q| + \sqrt{m^2-q^2} \leq r + \sqrt{m^2-q^2}.
\]
We note that the stability of the outermost horizon in fact implies $|q| \leq r$
provided the horizon is connected~\cites{gibbons,khuriweinsteinyamadaPRD}. In view of the above, the upper
bound in~\eqref{upper-lower-bound} is equivalent to~\eqref{charged-penrose-inequality} under the additional hypothesis
$|q|\leq r$. The proof of this latter statement will be based on an adaptation of Bray's 
conformal flow; see also~\cite{dainkhuriweinsteinyamada}.

We now introduce a few definitions and state our main theorem and a corollary. A
time-symmetric initial
data set $(M,g,E,B)$ consists of a $3$-manifold $M$, a Riemannian metric $g$, and vector fields $E$ and $B$.
We assume that the data satisfies the Maxwell constraints with no charges outside the horizon $\div_g E=\div_g
B=0$, and the charged DEC
\begin{equation} \label{dec}
  \mu=R_g-2(|E|^2+|B|^2)\geq0.
\end{equation}
We assume that
the data is SAF meaning that the complement of a compact set in $M$ is the finite
union of disjoint ends, and on each end the fields decay according to
\[
  g-\delta = O_2(|x|^{-1}), \quad E = O_1(|x|^{-2}), \quad B = O_1(|x|^{-2}),
\]
and $R_g$ is integrable. This guarantees that the ADM mass and the total electric and magnetic
charges
\begin{gather*}
  m = \frac1{16\pi} \int_{S_\infty} (g_{ij,j}-g_{jj,i}) \nu^i\, dA, \\
  q_E = \frac1{4\pi} \int_{S_\infty} E_i \nu^i\, dA, \quad
  q_B = \frac1{4\pi} \int_{S_\infty} B_i \nu^i\, dA,
\end{gather*}
are well defined. Here, $\nu$ is the outer unit normal, and the limit is taken in a designated end.
Without loss of generality, we assume that the magnetic charge $q_B=0$, and from now on denote $q=q_E$. This can
always be achieved by a fixed rotation in the $(E,B)$-space. Conformally compactifying all but the designated end, we
can now restrict our attention to surfaces which bound compact regions, and define $S_2$ to enclose $S_1$ to mean
$S_1=\D \Omega_1$, $S_2=\D \Omega_2$ and $\Omega_1\subset \Omega_2$. An \emph{outermost horizon} is a compact minimal
surface not enclosed
in any other compact minimal surface.

\begin{theorem} \label{main:theorem}
Let $(M,g,E,B)$ be a SAF initial data set satisfying the chargeless Maxwell constraints, 
the charged DEC, with ADM mass $m$, total
charge $q$, and admitting an outermost horizon of area $A=4\pi r^2$. Then the upper bound
in~\eqref{upper-lower-bound} holds with equality if and only if the data is RN. Suppose that $|q|\leq r$, then
\eqref{charged-penrose-inequality} holds with equality if and only if the data is RN.
\end{theorem}

As noted above the first statement follows from the second.

\begin{corollary}
Given $m$ and $q$, satisfying $m\geq |q|$, RN is the unique maximizer of $A$.
Given $A$ and $q$, satisfying $4\pi q^2 \leq A$, RN is the unique minimizer of $m$.
\end{corollary}

We point out that the hypothesis of no charges outside the horizon seems
necessary. On the one hand, our proof uses the divergence-free character of $E$ and $B$ 
in the final stage once we switch to 
inverse mean curvature flow. Indeed, we suspect that our conformal flow would not 
converge to \RN\ when charges are present outside the horizon. In fact, in~\cite{malec}, 
the authors conjecture that one could construct spherically symmetric counterexamples 
in this case.  On the other hand, the heuristic argument based on cosmic censorship would 
not apply since matter can carry charges out to infinity leading to a final state with a 
total charge different from the initial state. Thus, without additional hypotheses, one 
is not able to say how the upper bound in~\eqref{upper-lower-bound} for the 
final state compares to the same expression for the initial state.

In what follows, a brief outline of the main elements in the proof of 
Theorem~\ref{main:theorem} is given. Full details will appear in a forthcoming paper.

\section{The conformal flow}

Consider a SAF initial data set $(M_0,g_0,E_0,B_0)$ satisfying the Maxwell constraints and the charged DEC. We define
the
conformal flow
\begin{equation} \label{flow}
  g_t = u_t^4 g_0, \quad E^i_t=u_t^{-6} E^i_0, \quad B^i_t=u_t^{-6}B^i_0, \quad u_0=1.
\end{equation}
This immediately yields that the Maxwell constraints $\div_{g_t} E_t=\div_{g_t} B_t=0$ are preserved
under the flow and that the charge $q_t$ is constant. The logarithmic velocity of the flow $v_t=\dot u_t/u_t$ is
determined by
the following elliptic problem
\begin{equation} \label{velocity}
  \Delta_{g_t} v_t - (|E_t|^2+|B_t|^2) v_t=0,  \quad v_t\to-1 \> \text{at $\infty$},
  \quad v_t|_{\Sigma_t} = 0,
\end{equation}
where $\Sigma_t$ is the outermost horizon in $g_t$. We point out that by the maximum 
principle $-1<v<0$, and by the Hopf boundary Lemma, the outward normal 
derivative of $v$ on $\Sigma_t$ is negative. In particular, this guarantees that the 
surfaces $\Sigma_t$ always move outward.
Using the covariance $L_{g}(vu)=u^5L_{u^4g}v$ of the conformal Laplacian
$L_{g}=\Delta_{g}-\frac18 R_{g}$, we have
\[
  \frac18 \frac{d(u_{t}^{5}R_{g_{t}})}{dt} = -L_{g_0}\dot u_{t}
   =-u_{t}^{5}\left(|E_t|^2+|B_t|^2 - \frac18 R_{g_{t}}\right)v_{t},
\]
hence, from~\eqref{dec}, $u_t^4\mu_t$ is constant, and in particular $\mu_t\geq0$ for all $t$ provided $\mu_0\geq0$.
Thus the charged DEC is preserved. The
proof of the existence of solutions to~\eqref{flow}--\eqref{velocity} follows~\cite{bray2001} closely, and it is easily
checked that $A_t$ is constant.
The remaining two ingredients of the proof are to show that the
mass $m_t$ is non-increasing, and the inequality~\eqref{charged-penrose-inequality} holds at some final time
$T\in(0,\infty]$, implying that~\eqref{charged-penrose-inequality} holds also at the initial time $t=0$.

\section{Monotonicity} \label{monotonicity}

As in~\cite{bray2001} the proof of monotonicity of $m_t$ for our flow is based on a clever doubling argument by
Bunting--Masood-Ul-Alam first
introduced in~\cite{buntingmasood}. However here a more judicious choice of conformal factor, inspired
by~\cite{masood},
is required before we can apply the positive mass theorem. First, we note that since the
flow~\eqref{flow}--\eqref{velocity} is autonomous, it is enough to show that $\dot m_t\leq0$ at $t=0$. For
convenience we drop the subscript $0$.

We take two copies $M_{\pm}$ of the exterior of $\Sigma$, attach them at $\Sigma$, and equip them with conformal
metrics $g_{\pm}=w_{\pm}^4 g$, where $w_{\pm}=\frac{1}{2}\sqrt{(1\pm v)^2-\phi^2}$ and $\phi$ satisfies the differential inequality
\begin{equation} \label{phi}
  \phi\left(\Delta_g \phi-\frac{\nabla v\cdot\nabla\phi}{v}\right)
  \geq\Lambda \left||E|^{2}+|B|^2-\frac{|\nabla\phi|^{2}}{v^2}\right|,
\end{equation}
for some $\Lambda>0$ large enough, with boundary conditions $\partial_\nu\phi=0$ on $\Sigma$, $\phi\to0$ as
$|x|\to\infty$. From the asymptotic expansion, it turns out that $|x|\phi\to|q|$ at infinity.
Inequalities \eqref{dec} and \eqref{phi} guarantee that $R_{g_\pm}\geq0$ if $\Lambda\geq 12$, and the
boundary conditions guarantee that mean curvatures on both sides of the gluing agree. Furthermore the maximum principle,
$m\geq|q|$, and the asymptotics of $\phi$ guarantee that $(1\pm v)^2-\phi^2>0$, and the
asymptotics of $w_{\pm}$
guarantee that the $M_+$ end is compactified while the mass of the
$M_-$ end is given by $\widetilde m=m-\gamma$, where $\gamma$ is determined by 
$v=-1+\gamma/|x|+O(|x|^{-2})$. Since $v>-1$, we have $\gamma>0$. The
positive mass theorem~\cites{schoenyau1979,witten1981} can now be applied to conclude that $\widetilde m\geq0$ with
equality if and only if $(\widetilde M, g_{\pm})$ is the Euclidean space. Since, as in~\cite{bray2001} $\dot
m=2(\gamma-m)$ we get monotonicity with equality if and only if $(\widetilde M, g_{\pm})$ is flat.

It remains to show that~\eqref{phi} has a positive solution satisfying the required boundary conditions.
Since this part is very technical we leave the details to our forthcoming article. 
The main idea is to solve \eqref{phi} with equality replacing inequality on the exterior 
of a small neighborhood of the boundary
$\Omega=\{x\in M\mid \operatorname{dist}(x,\Sigma)>\tau\}$. We use the Leray-Schauder 
fixed point theorem~\cite{gilbargtrudinger}*{Theorem 11.6} to accomplish this, with 
appropriately chosen Dirchlet boundary conditions on $\partial\Omega$. Using such a domain 
avoids the difficulty of singular coefficients that occurs at $\Sigma$ due to the 
vanishing of $v$. Finally $\phi$ is then extended across $\partial\Omega$ while preserving 
the inequality \eqref{phi}. Although the regularity of the extended solution is only 
$C^{1,1}$ across
$\partial\Omega$, this is enough for an application of the positive mass theorem as described in the preceding paragraph.

\section{Exhaustion}

Considerable effort is spent in~\cite{bray2001} to show that the exterior of $\Sigma_t$ converges as
$t\to\infty$ to a Schwarzschild slice. We circumvent these difficulties and instead
obtain~\eqref{charged-penrose-inequality}
at a late time $T$. As in~\cite{bray2001}, we prove in two steps that the surface
$\Sigma_t$ eventually encloses any given compact surface. First, we show that
no compact surface in $M$ can enclose $\Sigma_t$ for all $t$. Then we
show that $\Sigma_t$ must eventually enclose any given compact surface.  It is here
that the hypothesis $|q|\leq r$ is used. Recall that this inequality is necesary for the connectedness of
the outermost horizon.  Thus at late times, $\Sigma_{T}$ is connected, and hence the inverse mean
curvature flow can be applied to obtain~\eqref{charged-penrose-inequality} for $(M_T,g_T,E_T,B_T)$, where $M_T$ is the
exterior of $\Sigma_T$.

After a perturbation, it may be assumed that the initial data set $(M,g,E,B)$ has charged
harmonic asymptotics~\cite{corvino}. That is, in the asymptotic end, $g=U_{0}^{4}\delta$,
$E=U_{0}^{-6}E_{\delta}$, $E_{\delta}=q\nabla r^{-1}$ where $\delta$ is the Euclidean metric,
$R_{g}=-8U_{0}^{-5}\Delta_{\delta}U_{0}=2|E|^{2}$, and $B=0$.

\begin{lemma}\label{exhaustion}
If $|q| < r$, then $\Sigma_{t}$ cannot be entirely enclosed by the coordinate sphere
$S_{r(t)}$ for all $t$, where $r(t)=\varepsilon re^{2t}$
for some sufficiently small $\varepsilon$.
\end{lemma}

Assume by contradiction that $\Sigma_{t}$ is entirely enclosed by $S_{r(t)}$ for all $t$.
We show that for some large $T$, $\Sigma_T$ is not
the outermost minimal area enclosure of $\Sigma_{0}$, yielding a contradiction.

Writing $U_{t}=u_{t}U_{0}$ and $V_{t}=v_{t}u_{t}U_{0}$, then
\[
 \Delta_{\delta}U_{t}=-\frac{1}{4}|E_{\delta}|^{2}U_{t}^{-3}, \quad
 \Delta_{\delta}V_{t}=\frac{3}{4}U_{t}^{-4}|E_{\delta}|^{2}V_{t}.
\]
Let $\V_{t}$ be the unique solution of the second equation above with $U_{t}$ replaced by $\U_{t}$, and
satisfying $\V_{t}=0$ on $S_{r(t)}$, and $\V_{t}\rightarrow -e^{-t}$ as $|x|\rightarrow \infty$,
where $\U_{t}$ is the conformal factor $U_{t}$ in the conformal flow of the Reissner-Nordstr\"{o}m initial data.
Note that $\V_{t}$ is the velocity $\tilde{v}_{t}\U_t$ in the conformal flow of the Reissner-Nordstr\"{o}m initial data,
where $\tilde{v}_t$ is obtained from~\eqref{RN-velocity} by setting
$m^2=4e^{-4t}r(t)^2+q^2$, and thus from~\eqref{U}
\[
\U_{t}=\left(e^{-2t}+\frac{\sqrt{4e^{-4t}r(t)^{2}+q^{2}}}{|x|}+e^{-2t}\frac{r(t)^{2}}{|x|^{2}}\right)^{1/2}.
\]

The idea is to compare $V_{t}$ and $\V_{t}$ to obtain estimates on $U_{t}$ in terms of $\U_{t}$.
However, we only need to estimate
$\int_{S_{r(t)}}U_{t}^{4}d\sigma_{\delta}$. Thus, let $\hatU_{t}$ be the unique solution of
\begin{gather*}
  \Delta_{\delta}\hatU_{t} = -\frac{1}{4}|E_{\delta}|^{2}\hatU_{t}^{-3}, \quad
    \hatU_{t}\to e^{-t}\> \text{as $|x|\to\infty$}, \\
  \hatU_{t}|_{S_{r(t)}}=\left(\frac{1}{4\pi r(t)^{2}} \int_{S_{r(t)}} U_{t}^{4}\right)^{1/4}.
\end{gather*}
This radial function can be computed explicitly
\begin{equation}
\hatU_{t}^{4}(x)=e^{-4t}+\frac{e^{-2t}\sqrt{\frac{8}{3}\left(\alpha+\frac{1}{2}q^{2}\right)}}{|x|}
+\frac{\alpha}{|x|^{2}}+\frac{e^{2t}\sqrt{\frac{8}{3}\left(\alpha+\frac{1}{2}q^{2}\right)}(\alpha-q^{2})}{6|x|^{3}}
+\frac{e^{4t}(\alpha-q^{2})^{2}}{36|x|^{4}},
\end{equation}
where $\alpha$ is a positive constant depending on $\int_{S_{r(t)}} U_{t}^4$. The assumption $|q|\leq r$
guarantees
that $\alpha\geq q^{2}+6e^{-4t}r(t)^{2}$, and hence
$\hatU_{t}(x)\geq\U_{t}(x)$ for $|x|\geq r(t)$.

Now $W_{t}=\V_{t}-\hatV_{t}$ satisfies
\[
  \Delta_{\delta}W_{t}=\frac{3}{4}\hatU_{t}^{-4}|E_{\delta}|^{2}W_{t}+\frac{3}{4}(\U_{t}^{-4}-\hatU_{t}^{-4}
  )\V_{t}|E_{\delta}|^{2},
\]
$W_{t}\rightarrow 0\> \text{as $|x|\rightarrow\infty$}$, and $W_{t}>0$ on $S_{r(t)}$ because
$\hatV_{t}(r(t))=\frac{d}{dt}\hatU_{t}(r(t))<0=\V_{t}(r(t))$. Therefore, since $\U_{t}^{-4}-\hatU_{t}^{-4}\geq 0$
the maximum principle gives that $W_{t}\geq 0$ outside $S_{r(t)}$.

This yields the upper bound $\hatV_{t}\leq\V_t$, and hence since $\hatV_{t}=\frac{d}{dt}\hatU_{t}$ it also gives an
estimate of $\hatU_{t}$ from above in terms of $\V_t$. This gives an upper bound on $\int_{S_{r(t)}} U_t^4$, and it then
follows as in~\cite{bray2001} that $|S_{r(t)}|\leq \varepsilon^{2}A[2+O(\varepsilon^{-1}e^{-t})]^{4}$.  Hence, for
$\varepsilon$ sufficiently small and $T$ sufficiently large, we have $|S_{r(T)}|<A$, and $\Sigma_T$ is not
outer area
minimizing, in contradiction to its definition.

\section{Rigidity}

In the case of equality, the mass $\widetilde m$ of the doubled manifold $(\widetilde M,g_\pm)$ in
the monotonicity proof must be zero, hence $\widetilde M$ is $\R^3$ and consequently $\Sigma$ is connected.
Thus, we can use Disconzi-Khuri's definition of the charged Hawking mass
\[
  m_{\text{CH}}(S) = \frac r2\left(1 - \frac{q^2}{r^2} - \frac1{16\pi} \int_S H^2\, dA \right),
\]
and its monotonicity under the inverse mean curvature to show that if equality holds when the horizon is connected, then
the initial data set is RN. Although $B$ is assumed to vanish in~\cite{disconzikhuri}*{Theorem 1},
the argument carries through in the time-symmetric case even if $B\ne0$.

Finally, we note that if the initial data set is RN, then the conformal flow defined by~\eqref{flow}
and~\eqref{velocity} simply yields a rescaling of RN as indeed it must by rigidity. The RN metric can be
written in isotropic coordinates as
$-V^{2}dt^{2}+g$, where $g=U^{4}\delta$,
\begin{equation} \label{U}
   U(x)=\left(1+\frac{m}{|x|}+\frac{m^{2}-q^{2}}{4|x|^{2}}\right)^{1/2},
\end{equation}
and the electric fields is $E_{i}=U^{-2}\partial_i(q/|x|)$. The conformal flow given by rescaling the coordinates
$x\mapsto e^{-2t}x$ has logarithmic flow velocity
\begin{equation} \label{RN-velocity}
  v_t= \frac{-e^{-2t}+e^{2t}(m^{2}-q^{2})/4|x|^{2}}{e^{-2t}+m/|x|+e^{2t}(m^{2}-q^{2})/4|x|^{2}}.
\end{equation}
It is now straightforward to verify that $v_t$ satisfies~\eqref{flow}--\eqref{velocity}.


\begin{bibdiv}
\begin{biblist}

\bib{bray2001}{article}{
      author={Bray, Hubert~L.},
       title={{Proof of the {R}iemannian {P}enrose inequality using the
  positive mass theorem}},
        date={2001},
        ISSN={0022-040X},
     journal={J. Differential Geom.},
      volume={59},
      number={2},
       pages={177\ndash 267},
         url={http://projecteuclid.org/getRecord?id=euclid.jdg/1090349428},
      review={\MR{1908823 (2004j:53046)}},
}

\bib{buntingmasood}{article}{
      author={Bunting, Gary~L.},
      author={Masood-ul Alam, A. K.~M.},
       title={{Nonexistence of multiple black holes in asymptotically
  {E}uclidean static vacuum space-time}},
        date={1987},
        ISSN={0001-7701},
     journal={Gen. Relativity Gravitation},
      volume={19},
      number={2},
       pages={147\ndash 154},
         url={http://dx.doi.org/10.1007/BF00770326},
      review={\MR{876598 (88e:83031)}},
}

\bib{corvino}{article}{
      author={Corvino, Justin},
       title={{On the asymptotics for the Einstein-Maxwell constraint
  equations}},
        date={2013},
     journal={In preparation},
}

\bib{dainkhuriweinsteinyamada}{article}{
      author={Dain, Sergio},
      author={Khuri, Marcus},
      author={Weinstein, Gilbert},
      author={Yamada, Sumio},
       title={{Lower Bounds for the Area of Black Holes in Terms of Mass,
  Charge, and Angular Momentum}},
        date={2013},
     journal={Phys.Rev.},
      volume={D88},
       pages={024048},
      eprint={1306.4739},
}

\bib{disconzikhuri}{article}{
      author={Disconzi, Marcelo~M.},
      author={Khuri, Marcus~A.},
       title={{On the {P}enrose inequality for charged black holes}},
        date={2012},
        ISSN={0264-9381},
     journal={Classical Quantum Gravity},
      volume={29},
      number={24},
       pages={245019, 18},
         url={http://dx.doi.org/10.1088/0264-9381/29/24/245019},
      review={\MR{3002957}},
}

\bib{geroch1973}{article}{
      author={Geroch, Robert},
       title={{Energy Extraction}},
        date={1973},
        ISSN={1749-6632},
     journal={Annals of the New York Academy of Sciences},
      volume={224},
      number={1},
       pages={108\ndash 117},
         url={http://dx.doi.org/10.1111/j.1749-6632.1973.tb41445.x},
}

\bib{gibbons}{article}{
      author={Gibbons, G.~W.},
       title={{Some comments on gravitational entropy and the inverse mean
  curvature flow}},
        date={1999},
        ISSN={0264-9381},
     journal={Classical Quantum Gravity},
      volume={16},
      number={6},
       pages={1677\ndash 1687},
         url={http://dx.doi.org/10.1088/0264-9381/16/6/302},
      review={\MR{1697098 (2000j:53089)}},
}

\bib{gibbonshawkinghorowitzperry}{article}{
      author={Gibbons, G.~W.},
      author={Hawking, S.~W.},
      author={Horowitz, Gary~T.},
      author={Perry, Malcolm~J.},
       title={{Positive mass theorems for black holes}},
        date={1983},
        ISSN={0010-3616},
     journal={Comm. Math. Phys.},
      volume={88},
      number={3},
       pages={295\ndash 308},
         url={http://projecteuclid.org/getRecord?id=euclid.cmp/1103922377},
      review={\MR{701918 (84k:83015)}},
}

\bib{gilbargtrudinger}{book}{
      author={Gilbarg, David},
      author={Trudinger, Neil~S.},
       title={{Elliptic partial differential equations of second order}},
      series={{Classics in Mathematics}},
   publisher={Springer-Verlag},
     address={Berlin},
        date={2001},
        ISBN={3-540-41160-7},
        note={Reprint of the 1998 edition},
      review={\MR{1814364 (2001k:35004)}},
}

\bib{hawkingellis}{book}{
      author={Hawking, S.~W.},
      author={Ellis, G. F.~R.},
       title={{The large scale structure of space-time}},
   publisher={Cambridge University Press},
     address={London},
        date={1973},
        note={Cambridge Monographs on Mathematical Physics, No. 1},
      review={\MR{0424186 (54 \#12154)}},
}

\bib{huiskenilmanen2001}{article}{
      author={Huisken, Gerhard},
      author={Ilmanen, Tom},
       title={{The inverse mean curvature flow and the {R}iemannian {P}enrose
  inequality}},
        date={2001},
        ISSN={0022-040X},
     journal={J. Differential Geom.},
      volume={59},
      number={3},
       pages={353\ndash 437},
         url={http://projecteuclid.org/getRecord?id=euclid.jdg/1090349447},
      review={\MR{1916951 (2003h:53091)}},
}

\bib{jang1979}{article}{
      author={Jang, Pong~Soo},
       title={{Note on cosmic censorship}},
        date={1979},
     journal={Phys.Rev.},
      volume={D20},
       pages={834\ndash 838},
}

\bib{waldjang}{article}{
      author={Jang, Pong~Soo},
      author={Wald, Robert~M.},
       title={{The positive energy conjecture and the cosmic censor
  hypothesis}},
        date={1977},
        ISSN={0022-2488},
     journal={J. Mathematical Phys.},
      volume={18},
      number={1},
       pages={41\ndash 44},
      review={\MR{0523907 (58 \#25755)}},
}

\bib{khuriweinsteinyamadaPRD}{article}{
      author={Khuri, Marcus~A},
      author={Yamada, Sumio},
      author={Weinstein, Gilbert},
       title={{On the Riemannian Penrose inequality with charge and the cosmic
  censorship conjecture}},
        date={2012},
     journal={Res. Inst. Math. Sci. Kokyuroku},
      number={1862},
       pages={63\ndash 66},
      eprint={1306.0206},
}

\bib{malec}{article}{
      author={Malec, Edward},
      author={Murchadha, Niall~{\'O}},
       title={{Trapped surfaces and the Penrose inequality in spherically
  symmetric geometries}},
        date={1994Jun},
     journal={Phys. Rev. D},
      volume={49},
       pages={6931\ndash 6934},
         url={http://link.aps.org/doi/10.1103/PhysRevD.49.6931},
}

\bib{masood}{article}{
      author={Masood-ul Alam, A. K.~M.},
       title={{Uniqueness proof of static charged black holes revisited}},
        date={1992},
        ISSN={0264-9381},
     journal={Classical Quantum Gravity},
      volume={9},
      number={5},
       pages={L53\ndash L55},
         url={http://stacks.iop.org/0264-9381/9/L53},
      review={\MR{1163879 (93f:83059)}},
}

\bib{penrose1982}{incollection}{
      author={Penrose, R.},
       title={{Some unsolved problems in classical general relativity}},
        date={1982},
   booktitle={{Seminar on {D}ifferential {G}eometry}},
      series={{Ann. of Math. Stud.}},
      volume={102},
   publisher={Princeton Univ. Press},
     address={Princeton, N.J.},
       pages={631\ndash 668},
      review={\MR{645761 (83c:83001)}},
}

\bib{penrose1973}{article}{
      author={Penrose, Roger},
       title={{Naked Singularities}},
        date={1973},
        ISSN={1749-6632},
     journal={Annals of the New York Academy of Sciences},
      volume={224},
      number={1},
       pages={125\ndash 134},
         url={http://dx.doi.org/10.1111/j.1749-6632.1973.tb41447.x},
}

\bib{schoenyau1979}{article}{
      author={Schoen, {R}ichard},
      author={Yau, {S}hing~{T}ung},
       title={{On the proof of the positive mass conjecture in general
  relativity}},
        date={1979},
        ISSN={0010-3616},
     journal={Comm. Math. Phys.},
      volume={65},
      number={1},
       pages={45\ndash 76},
         url={http://projecteuclid.org/getRecord?id=euclid.cmp/1103904790},
      review={\MR{526976 (80j:83024)}},
}

\bib{weinsteinyamada}{article}{
      author={Weinstein, Gilbert},
      author={Yamada, Sumio},
       title={{On a {P}enrose inequality with charge}},
        date={2005},
        ISSN={0010-3616},
     journal={Comm. Math. Phys.},
      volume={257},
      number={3},
       pages={703\ndash 723},
         url={http://dx.doi.org/10.1007/s00220-005-1355-0},
      review={\MR{2164949 (2007c:83016)}},
}

\bib{witten1981}{article}{
      author={Witten, Edward},
       title={{A new proof of the positive energy theorem}},
        date={1981},
        ISSN={0010-3616},
     journal={Comm. Math. Phys.},
      volume={80},
      number={3},
       pages={381\ndash 402},
         url={http://projecteuclid.org/getRecord?id=euclid.cmp/1103919981},
      review={\MR{626707 (83e:83035)}},
}

\end{biblist}
\end{bibdiv}
\bibliographystyle{}


\end{document}